\documentclass[english,aps,preprint]{revtex4}
\usepackage[]{fontenc}
\usepackage[latin9]{inputenc}
\usepackage{graphicx}
\usepackage{amssymb}
\usepackage{babel}

\begin{document}

\preprint{Submitted to Appl. Phys. Lett.}

\title{Charge fluctuations for particles on a surface exposed to plasma}

\author{T. E. Sheridan}

\email{t-sheridan@onu.edu}

\address{Department of Physics \& Astronomy, Ohio Northern University, Ada,
Ohio 45810}

\author{A. Hayes}

\address{Veeco Process Equipment, Terminal Dr., Plainview, New York 11803}

\begin{abstract}
We develop a stochastic model for the charge fluctuations on a microscopic
dust particle resting on a surface exposed to plasma. We find in steady
state that the fluctuations are normally distributed with a standard
deviation that is proportional to $\left(CT_{e}\right)^{1/2}$, where
$C$ is the particle-surface capacitance and $T_{e}$ is the plasma
electron temperature. The time for an initially uncharged ensemble
of particles to reach the steady state distribution is directly proportional
to $CT_{e}$. 
\end{abstract}


\maketitle
As features on integrated circuits become smaller, the problem of
contamination by microscopic particles becomes larger. The best solution
is to minimize particle creation and to prevent particles from reaching
the substrate. However, if particles are deposited on the substrate
then methods for removing them are needed. One proposed method is
called plasma-assisted electrostatic cleaning (PAEC). 

\begin{figure}
\includegraphics[width=3.25in]{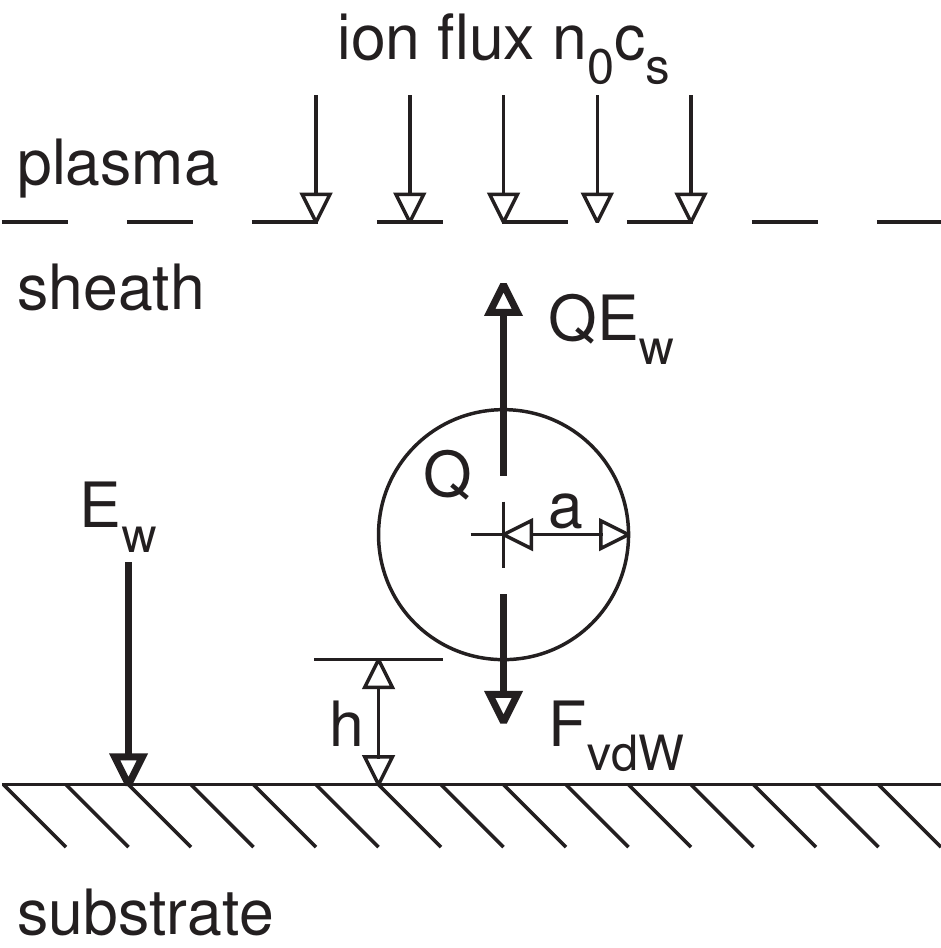}

\caption{\label{fig:paec}Schematic of plasma assisted electrostatic cleaning
(PAEC). A particle on a surface exposed to plasma acquires a charge
$Q<0$ and is pulled from the surface when the electrostatic force
$QE_{w}$ due to the sheath electric field $E_{w}$ exceeds the adhesive
van der Waals force $F_{{\rm vdW}}$.}

\end{figure}

PAEC works by exposing a surface with dust particles on it to a plasma
(Fig. \ref{fig:paec}). It is conjectured \cite{tes2} that a particle
can acquire a net charge $Q<0$ from the fluxes of charged plasma
species. The electrostatic force on the particle due to the sheath
electric field $E_{w}$ is $QE_{w}$. When this force exceeds the
adhesive force the particle is pulled from the surface and experiences
a large acceleration across the sheath. If there are no nearby regions
where particles can be trapped \cite{wan2,wan1}, the particle will
move away, and the surface will have been {}``cleaned.'' Thus, the
plasma serves two purposes: charging the contamination particle and
enhancing the electric field at the substrate. Several experiments
\cite{tes2,flan,lyt} have shown that plasma exposure can remove micrometer
to nanometer-diameter particles from both insulating and conducting
surfaces.

Sheridan and Goree \cite{tes2} coated an aluminum surface with 0.2--10
$\mu$m alumina grains and found that significant particle removal
occurred in a low-density plasma ($n_{e}\approx10^{8}{\rm cm}^{-3}$)
when the surface was exposed to energetic electrons (59 eV) from an
emissive filament. They observed that the removal rate increased with
plasma density and decayed exponentially with time. Flanagan and Goree
\cite{flan} used JSC-1 (Johnson Space Center lunar simulant one)
particles on a glass substrate, again for low plasma densities and
with energetic electrons, and verified that the time constant for
removal decreases with increasing plasma density. They found time
constants ranging from $30\,{\rm s}$ to $1\,{\rm s}$ for plasma
densities from $2\times10^{6}\;{\rm cm}^{-3}$ to $2\times10^{8}\;{\rm cm}^{-3}$.
Lytle et al. \cite{lyt} performed experiments using a high density
pulsed helicon source to remove particles as small as 30 nm from a
dielectric substrate. Their results demonstrate that a minimum plasma
exposure time is required for particle removal. In Lytle's studies
the reported electron temperature was low ($\approx3{\rm ~eV}$),
but the plasma density was 3 to 4 orders of magnitude larger than
that in the other experiments \cite{tes2,flan}.

The dominant force holding a microscopic particle to a surface is
thought to be the London-van der Waals force \cite{got,li}. The adhesive
force on 10--100-nm diameter particles is reported \cite{li} to be
$\sim1$--10 nN. An estimate for the van der Waals force between a
spherical particle of radius $a$ and a smooth, flat surface is \cite{flan}
$F_{{\rm vdW}}=Ha/(6h^{2})$, where the Hamaker constant $H\sim10^{-19}\;{\rm J}$
and $h$ is the separation between the particle and the surface (Fig.
\ref{fig:paec}). For a characteristic separation $h=0.3\;{\rm nm}$
and radius $a=10\;{\rm nm}$, $F_{{\rm vdW}}\approx2\;{\rm nN}$.
Flanagan \cite{flan} estimated for his experiment that $E_{w}$ may
have been as large as 10 kV/m . For this value of $E_{w}$ a particle
charge $Q\sim-10^{6}e$ would be required to overcome the predicted
adhesive force. There is no evidence that microscopic particles can
attain such large charges; the potential of a 10-nm radius particle
in free space with this charge would be $-150\,{\rm kV}$. Consequently,
it seems likely that adhesive forces are significantly overestimated
for typical processing surface conditions and that most particles
removed by plasma cleaning are loosely bound. This may be due to surface
contamination and/or to the fact that most particles are not perfect
spheres and so rest on the surface at a small number of contact points
whose asperity size is much less than the particle size, giving a
reduced effective radius.

One way that a particle on a biased surface can acquire charge is
by sharing charge with that surface. Flanagan \cite{flan} referred
to this as the {}``shared charge model.'' The shared charge on a
spherical particle on a flat surface with electric field $E_{w}$
is \cite{xwan} $Q=(1.64)4\pi\epsilon_{0}E_{w}a^{2}$, where the factor
of 1.64 is a small enhancement because the particle is a perturbation
to the otherwise flat surface. For microscopic particles under typical
plasma conditions, the shared charge is small. For example, for $E_{w}=10\,{\rm kV/m}$
and $a=500\,{\rm nm}$, $Q=-2.9e$. That is, on average, microscopic
particles on a surface have $\left|Q\right|\lesssim e$ due to shared
charge. Such an average charge is much too small to explain plasma
assisted cleaning. 

Particles can also acquire charge because of electron and ion fluxes
from the plasma. In steady state, these fluxes balance. However, fluctuations
in the particle charge may be significant due to the discrete nature
of the charging process. It is found experimentally \cite{tes2,flan}
that the rate of particle release during plasma cleaning decays exponentially,
indicating that each particle release is an independent random event.
This is consistent with release being due to particle charge fluctuations,
where release occurs during large negative excursions in charge. In
what follows, we develop a model of charge fluctuations caused by
electron and ion fluxes to a particle on a surface.

The ion flux across the sheath edge is $\Gamma_{i}=n_{0}c_{s}$ (Fig.
\ref{fig:paec}) where $n_{0}$ is the plasma density at the sheath
edge, $c_{s}=\sqrt{eT_{e}/m_{i}}$ is the ion acoustic speed and $T_{e}$
is the effective electron temperature in eV. We assume a single ion
species with charge $+e$, and neglect secondary electron emission.
Since the sheath is essentially source-free, continuity requires that
$\Gamma_{i}$ is also the ion flux at the substrate. The frequency
with which ions strike a particle on the surface with ion collection
area $A_{i}$ is $\nu_{i}=\Gamma_{i}A_{i}$. The probability that
a particle collects an ion in a time interval $\Delta t\ll\nu_{i}^{-1}$
is then $P_{i}=\nu_{i}\Delta t$, which is assumed independent of
the particle's potential. 

For electrons in thermal equilibrium (i.e., Boltzmann electrons),
the electron flux to a surface at a potential $\phi$ with respect
to the sheath edge is\begin{equation}
\Gamma_{e}(\phi)=\frac{1}{4}n_{0}\bar{v}e^{\phi/T_{e}},\label{eq:Ge}\end{equation}
where $\bar{v}$ is the average electron speed. The average frequency
with which electrons strike a particle with electron collecting area
$A_{e}$ is then $\nu_{e}=\Gamma_{e}A_{e}$, and the probability of
collecting an electron in $\Delta t$ is $P_{e}\left(\phi\right)=\nu_{e}\Delta t$.
In the typical case where $\Gamma_{i}<\Gamma_{e}(0)$ the particle
will have a steady-state floating potential $\phi_{f}<0$ such that
the electron and ion collection rates are, on average, equal. Assuming
equal collection areas, $A_{i}=A_{e}$, $\phi_{f}$ is a solution
of $\nu_{i}=\nu_{e}\left(\phi_{f}\right)$. 

Now consider a fluctuation around the floating potential, where the
particle and surface are assumed to have the same average $\phi_{f}$.
The particle's potential can be written as $\phi=\phi_{f}+\delta\phi$,
where $\delta\phi$ is the potential difference between the particle
and the surface. The probability of electron collection then becomes\begin{equation}
P_{e}=P_{i}e^{\delta\phi/T_{e}},\label{eq:Pedphi}\end{equation}
which decreases when $\delta\phi<0$ because electrons are repelled,
and increases when $\delta\phi>0$ because electrons are attracted.
This mechanism effectively limits the particle's charge fluctuations
to a finite range about an average value $\delta\phi_{{\rm ave}}=0$. 

Potential fluctuations can be related to charge fluctuations $\delta Q$
by $\delta\phi=\delta Q/C$, where $C$ is the capacitance of the
particle-surface system. The probability of electron collection is
then \begin{equation}
P_{e}=P_{i}e^{\delta Q/(CT_{e})}.\label{eq:PedQ}\end{equation}
The charge on each particle performs a bounded random walk where the
probability of gaining an ion (a step to the {}``right'') is constant,
and the probability of gaining an electron (a step to the {}``left'')
is given by Eq. (\ref{eq:PedQ}). The charging model {[}Eq. (\ref{eq:PedQ})]
represents a Markov process \cite{kolm} for transition probabilities
from a charge state $\delta Q_{j}=je$, $j\in\left(-\infty,\infty\right)$.
Since the Markov process is regular and effectively finite the steady-state
distribution of $\delta Q$ is a normal (i.e., Gaussian) distribution.
This is our first result.

Our second result is that the steady state, root-mean-squared value
of the charge fluctuation (i.e., the standard deviation of the $\delta Q$
distribution) is \begin{equation}
\frac{\delta Q_{rms}}{e}=\sqrt{\frac{CT_{e}}{e}},\label{eq:sj}\end{equation}
which gives the fluctuations in units of $e$. Consequently, $\delta Q_{rms}$
depends only on $CT_{e}$. Note that $\delta Q_{rms}$ is \emph{not}
given by Poisson statistics for the net number of elementary charges
on the dust particle \cite{cui}, since in our model $\delta Q_{{\rm ave}}\ll\delta Q_{{\rm rms}}$.
We also find that the steady-state value of $\delta Q_{{\rm rms}}$
is independent of the plasma density, as is the case for isolated
particles in plasma \cite{jag1}. To further characterize the charge
fluctuations, we must consider $C$ .

The capacitance of an isolated spherical particle with radius $a$
is $C_{0}=4\pi\epsilon_{0}a$, which is valid in plasma when $a$
is small compared to the Debye length. If the particle is placed a
height $h\lesssim0.1a$ above a flat conducting plate (Fig. \ref{fig:paec}),
the capacitance increases to \cite{boy}\begin{equation}
\frac{C}{C_{0}}\approx\gamma+\frac{1}{2}\ln2-\frac{1}{2}\ln\frac{h}{a}>1,\label{eq:Cclose}\end{equation}
where $\gamma=0.5772\ldots$ is the Euler-Masherone constant. Here
$C$ diverges logarithmically as $h/a\rightarrow0$, so that $C>C_{0}$,
increasing the range of the charge fluctuations {[}Eq. (\ref{eq:sj})].
The capacitance is insensitive to the shape of compact particles when
$a$ is taken to be the characteristic particle size \cite{jag1}.
For values typical of Flanagan's experiment \cite{flan}, $a=500{\rm ~nm}$,
$h=0.3{\rm ~nm}$, and $T_{e}=60{\rm ~eV}$ (the primary electron
energy), we find $C/C_{0}=4.63$ and $\delta Q_{{\rm rms}}=311e$,
which greatly exceeds the shared charge.

To determine the temporal behavior of $\delta Q$, the model was solved
using a Monte Carlo simulation for an ensemble of $n$ identical particles.
For each particle, we start with the initial condition $\delta Q=0$.
For each time step $\Delta t$, we compute two random numbers distributed
uniformly in $[0,1)$. If the first random number is less than $P_{i}$,
then $\delta Q$ is increased by one, representing ion collection.
If the second random number is less than $P_{e}$, then $\delta Q$
is decreased by one, represented electron collection. 

\begin{figure}
\includegraphics[width=3.25in]{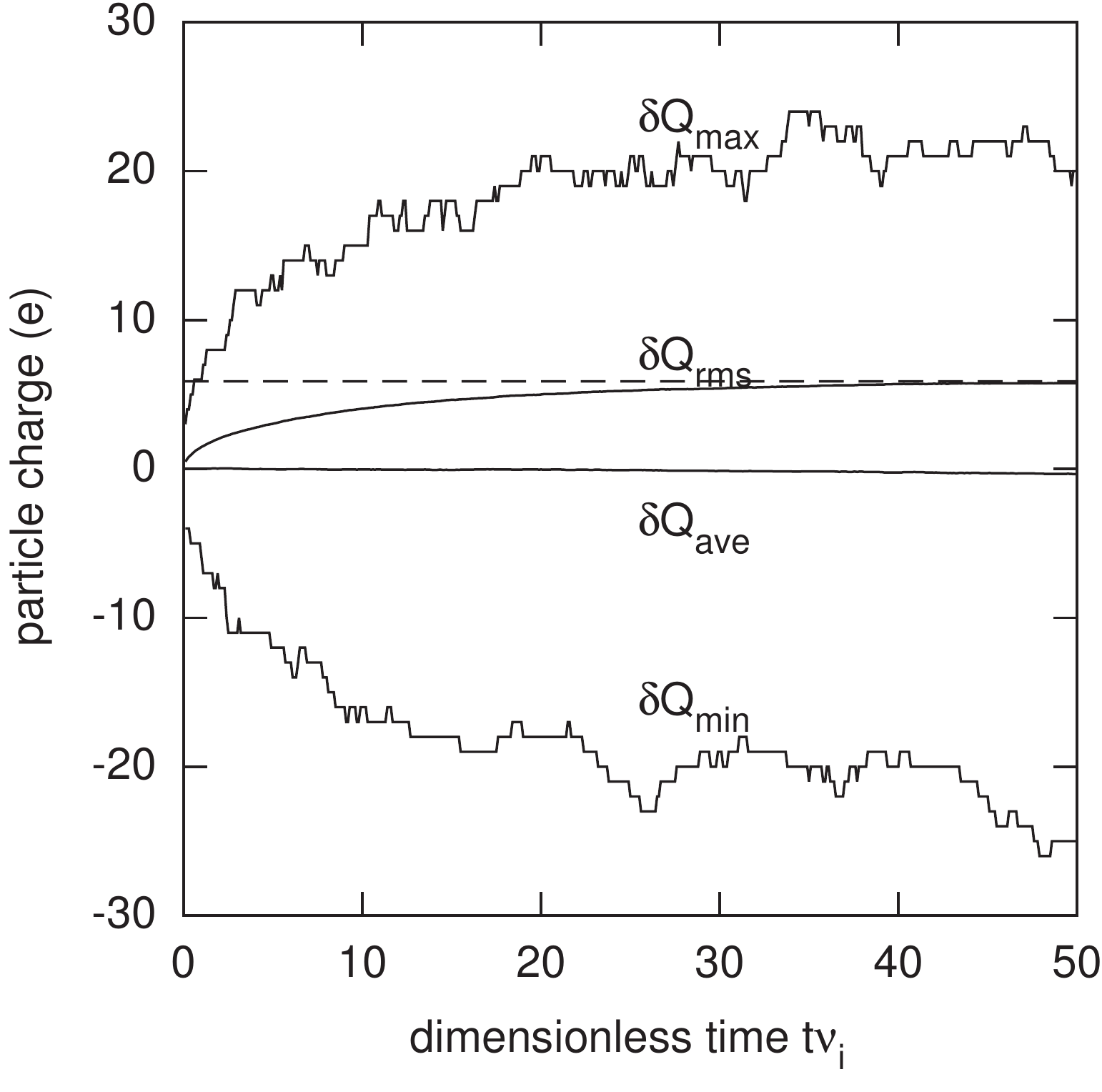}

\caption{\label{fig:sigQ}Time dependence of charge for an ensemble of $n=10\,000$
particles with $CT_{e}=5.56\times10^{-18}{\rm F\, eV}$. The dashed
line gives the asymptotic dependence of Eq. (\ref{eq:sj}).}

\end{figure}

For a given value of $CT_{e}$, we simulated this system for $n=10\,000$
particles vs dimensionless time $t\nu_{i}$. During the course of
the simulation the average charge $\delta Q_{ave}$, the standard
deviation $\delta Q_{rms}$, the largest positive charge $\delta Q_{max}$,
and largest negative charge $\delta Q_{min}$ were computed for the
ensemble. A time history of one simulation run is shown in Fig. \ref{fig:sigQ}
for $CT_{e}=5.56\times10^{-18}{\rm F\, eV}$, which corresponds to
the capacitance of an isolated sphere with $a=20\;{\rm nm}$ for $T_{e}=1\;{\rm eV}$.
Here the average of the charge fluctuations is approximately zero,
while the extreme values make excursions of up to $\pm25e$ away from
zero. When a particle's charge makes a large negative excursion, the
electrostatic force may exceed the adhesive force leading to particle
release.

\begin{figure}
\includegraphics[width=3.25in]{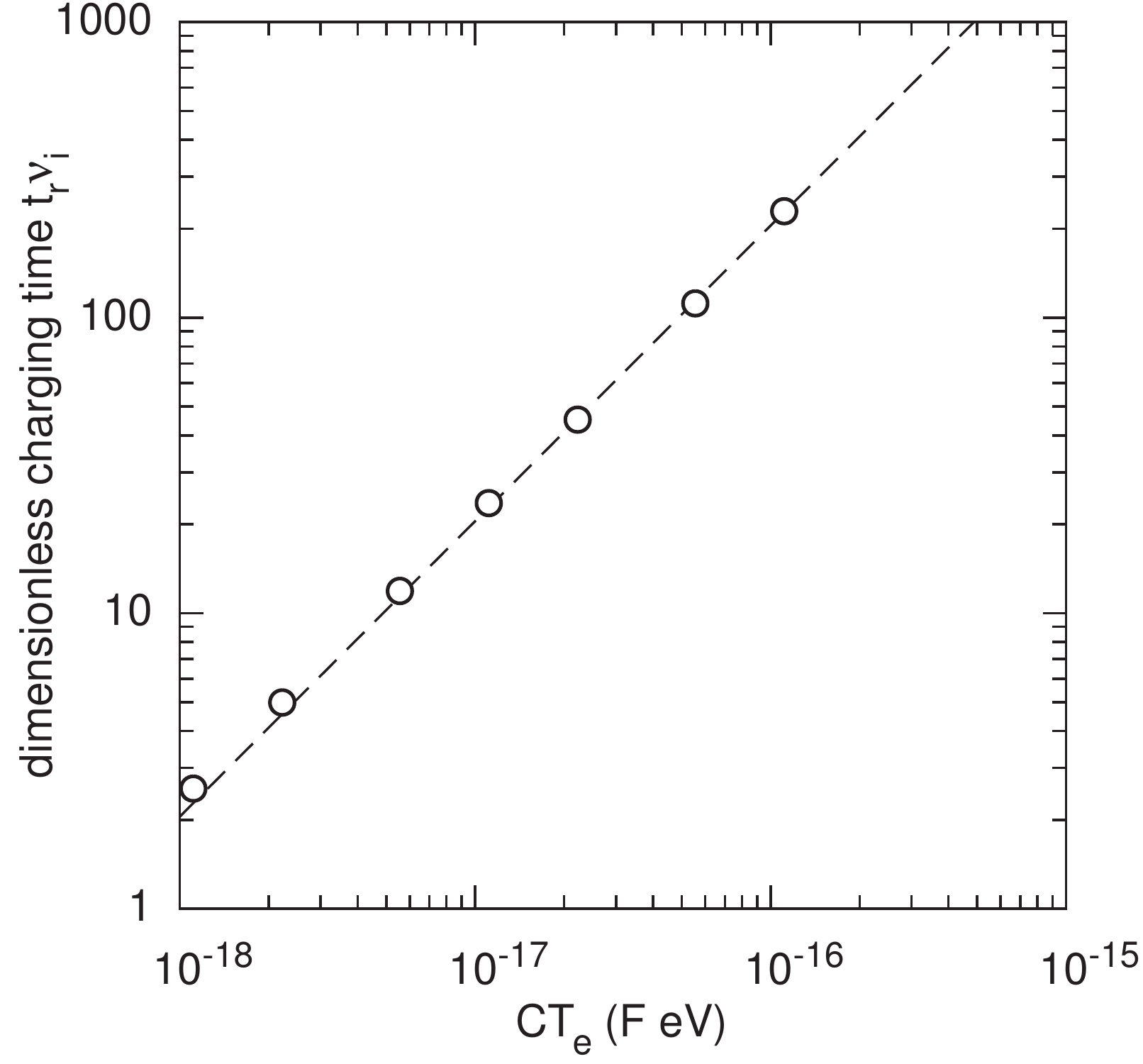}

\caption{\label{fig:sigQvsT}Dimensionless charging time $t_{r}\nu_{i}$ vs
particle-surface capacitance $C$ times the electron temperature $T_{e}$
in eV, determined from Monte Carlo simulations. The data are well
fitted by a straight line.}

\end{figure}

The dependence of the dimensionless charging time $t_{r}\nu_{i}$
on $CT_{e}$ computed using Monte Carlo simulations is shown in Fig.
\ref{fig:sigQvsT}. Here $t_{r}$ is the characteristic time required
for an initially uncharged ensemble to approach steady-state. The
charging time was found by fitting $\delta Q_{{\rm rms}}\left(t\right)$
with a function $\propto\left(1-e^{-t/t_{r}}\right)$. Our third result
is that the charging time increases linearly with $CT_{e}$ and is
well described by the line\begin{equation}
t_{r}\nu_{i}=\left(2.05\times10^{18}{\rm F}^{-1}{\rm eV}^{-1}\right)CT_{e}.\label{eq:tr}\end{equation}
The charging time in seconds is\begin{equation}
t_{r}=\left(72.6\times10^{6}{\rm m}^{-1}{\rm eV}^{-1}\right)\frac{C/C_{0}}{an_{0}c_{s}}T_{e}\propto\frac{1}{an_{0}},\label{eq:trdim}\end{equation}
where $a$ is in m, $n_{0}$ is in m$^{-3}$ and $c_{s}$ is in m/s.
Here $t_{r}$ is inversely proportional to the particle radius and
the plasma density. The inverse dependence on plasma density is consistent
with the measured \cite{flan} dependence of the particle release
time constant. Smaller particles have a longer charging time since
they have smaller collecting areas.

Finally, we can roughly estimate the electrostatic force for a dust
particle on an electrically floating surface in a discharge with a
bulk component and a low-density energetic electron component \cite{tes2,flan}.
For an electrically floating (or dielectric) substrate, the electric
field at the surface scales as $E_{w}\sim T_{e,hot}/\lambda_{D}$,
where $T_{e,hot}$ is the effective temperature of the energetic electrons
and determines the floating potential. For a low-pressure filament
discharge, $T_{e,hot}$ is roughly the primary electron energy. The
Debye shielding length $\lambda_{D}\sim T_{e,bulk}^{1/2}n_{0}^{-1/2}$
is dominated by the bulk electron space charge. Consequently, we predict
the electrostatic cleaning force scales as \begin{equation}
F\propto\delta Q_{rms}E_{w}\sim a^{1/2}n_{0}^{1/2}T_{e,hot}^{3/2}T_{e,bulk}^{-1/2},\label{eq:paecscale}\end{equation}
which gives a weak scaling with density and radius and a stronger
scaling with the the tail electron energy. This scaling is supported
by the experimental observations \cite{tes2,flan} that particle removal
occurs even at low plasma density when there is an energetic electron
component. Since the adhesive van der Waals force $\propto a$, it
goes to zero faster than the the electrostatic removal force {[}Eq.
(\ref{eq:paecscale})]. Consequently, there should be no lower limit
on the size of particles that can be removed using plasma assisted
cleaning.

\begin{acknowledgments}
The authors would like to acknowledge valuable comments from T. M.
Flanagan.
\end{acknowledgments}


\begin{thebibliography}{10}
\bibitem{tes2}T. E. Sheridan and J. Goree, J. Geophys. Res. \textbf{97},
2935--2942 (1992).

\bibitem{wan2}H. H. Hwang and M. J. Kushner, Appl. Phys. Lett. \textbf{68},
3716--3718 (1996).

\bibitem{wan1}H. H. Hwang, E. R. Keiter and M. J. Kushner, J. Vac.
Sci. Technol. \textbf{A 16}, 2454--2462 (1998).

\bibitem{flan}T. M. Flanagan and J. Goree, Phys. Plasmas \textbf{13},
123504 (2006).

\bibitem{lyt}W. M. Lytle, H. Shin and D. N. Ruzic, Proc. SPIE \textbf{6518},
65183P (2007).

\bibitem{got}M. G\"otzinger and W. Peukert, Langmuir \textbf{20},
5298--5303 (2004).

\bibitem{li}Q. Li, V. Rudolph and W. Peukert, Powder Tech. \textbf{161},
248--255 (2006).

\bibitem{xwan}X. Wang, J. Colwell, M. Horanyi and S. Robertson, IEEE
Trans. Plasma Sci. \textbf{35}, 271--279 (2007).

\bibitem{cui}C. Cui and J. Goree, IEEE Trans. Plasma Sci. \textbf{22},
151--158 (1994).

\bibitem{kolm}B. Kolman, \emph{Introductory Linear Algebra with Applications},
2nd ed. (Macmillan, New York, 1980) p. 433.

\bibitem{jag1}J. Goree, Plasma Sources Sci. Technol. \textbf{3},
400--406 (1994).

\bibitem{boy}L. Boyer, F. Houz\'e, A. Tonck, J.-L. Loubet and J.-M.
Georges, J. Phys. D: Appl. Phys. \textbf{27}, 1504--1508 (1994).
\end{thebibliography}
\end{document}